\documentclass[useAMS,usenatbib]{mn2e}

\usepackage{txfonts,graphicx,amssymb,subfigure}
\usepackage[normalem]{ulem}
\usepackage{color}

\newcommand{\Brpa}{\,B_{\parallel}}

\newcommand{\pdt}[1][long]{\ifthenelse{\equal{#1}{long}}
  {\frac{\partial}{\partial t}}{\frac{\partial #1}{\partial t}}}
\newcommand{\pdr}[1][long]{\ifthenelse{\equal{#1}{long}}
  {\frac{\partial}{\partial r}}{\frac{\partial #1}{\partial r}}}
\newcommand{\pddr}[1][long]{\ifthenelse{\equal{#1}{long}}
  {\frac{\partial^2}{\partial r^2}}{\frac{\partial^2 #1}{\partial r^2}}}
\newcommand{\pdp}[1][long]{\ifthenelse{\equal{#1}{long}}
  {\frac{\partial}{\partial \phi}}{\frac{\partial #1}{\partial \phi}}}
\newcommand{\pddp}[1][long]{\ifthenelse{\equal{#1}{long}}
  {\frac{\partial^2}{\partial \phi^2}}{\frac{\partial^2 #1}{\partial \phi^2}}}
\newcommand{\pdz}[1][long]{\ifthenelse{\equal{#1}{long}}
  {\frac{\partial}{\partial z}}{\frac{\partial #1}{\partial z}}}
\newcommand{\pddz}[1][long]{\ifthenelse{\equal{#1}{long}}
  {\frac{\partial^2}{\partial z^2}}{\frac{\partial^2 #1}{\partial z^2}}}

\title{Wavelet-based Faraday Rotation Measure Synthesis}

\author[P.~Frick et al.]
       {P.~Frick$^1$, D.~Sokoloff$\,^{2}$,  R.~Stepanov$^1$,
        and R.~Beck$^3$\\
$^1$ Institute of Continuous Media Mechanics,
Korolyov str.~1, 614013 Perm, Russia \\
$^2$ Department of Physics, Moscow University, 119899, Moscow, Russia \\
$^3$ Max-Planck-Institut f\"ur Radioastronomie, Auf dem H\"ugel 69,
  53121 Bonn, Germany}

\date{Accepted 2009 .... Received 2009 ....; in original form 2009}

\pagerange{\pageref{firstpage}--\pageref{lastpage}}
\pubyear{2009}
\begin{document}
\maketitle

\label{firstpage}

%
\begin{abstract}
Faraday Rotation Measure (RM) Synthesis, as a method for
analyzing multi-channel observations of polarized radio emission to
investigate galactic magnetic fields structures, requires the
definition of complex polarized intensity in the wavelength range
$-\infty <\lambda^2 < \infty$. The problem
is that the measurements at negative $\lambda^2$ are not possible. We introduce a simple method for continuation of
the observed complex polarized intensity $P(\lambda^2)$ into the
domain  $\lambda^2<0$ using symmetry arguments. The method is suggested in context of
magnetic field recognition in galactic disks where the magnetic field is supposed to have a maximum in
the equatorial plane. The  method is quite simple
when applied to a single Faraday-rotating structure on
the line of sight. Recognition of several structures on the same
line of sight requires a more sophisticated technique.  We
also introduce a wavelet-based algorithm which allows us to
consider a set of isolated structures in the
($\phi,\lambda^2$) plane (where $\phi$ is the Faraday depth).
The method essentially improves the possibilities for
reconstruction of complicated Faraday structures using the
capabilities of modern radio telescopes.
\end{abstract}


\section{Introduction}

Observations of polarized radio emission are the main
sources of information on magnetic fields of galaxies.
The basic idea of magnetic field analysis from
polarized radio emission data originates in the classical
paper of \cite{Burn1966MNRAS.133...67B} (for a later development see
\cite{1998MNRAS.299..189S}). In particular,
\cite{Burn1966MNRAS.133...67B} noted that the complex polarized
intensity $P$ obtained from a radio source is related to the Faraday
dispersion function $F(\phi)$  as
\begin{equation}
\label{p_to_f}
P(\lambda^2) =  \int_{-\infty}^{\infty} F(\phi) e^{2i\phi \lambda^2}  d \phi.
\end{equation}
$F(\phi)$ is the fraction of radiation with the Faraday depth $\phi$ multiplied by intrinsic complex polarization and it is an important emission characteristic of interest.
Here the Faraday depth $\phi$ is defined by
\begin{equation}
\phi(z) = -0.81\int_{z}^{0} \Brpa n_e dz',
\label{fardep}
\end{equation}
where $\Brpa$ is the line-of-sight magnetic field component measured
in $\mu$G, $n_e$ is the thermal electron density measured in
cm$^{-3}$ and the integral is taken from the
observer at $z=0$ over the region which
contains both, magnetic fields and free electrons,
and $z$ is measured in parsecs. Following
Eq.~(\ref{p_to_f}) $P$ is the inverse Fourier transform of $F$.
Correspondingly, the Faraday dispersion function $F$ is the Fourier
transform of the complex polarized intensity:
\begin{equation}
\label{f_to_p1}
F(\phi) = {{1} \over{\pi}} \hat P(k),
\label{Burn}
\end{equation}
where $k=2\phi$, and the Fourier transform is defined as
\begin{equation}
\label{four}
f\left( x \right) = {{1}\over{2\pi}} \int_{-\infty}^{\infty} {\hat {f}\left( k \right)e^{i k x}d k}, \quad
\hat {f}\left( k \right) = \int_{-\infty}^{\infty} {f\left( x \right)e^{ - i k x} dx}.
\end{equation}

\begin{figure}\centering{
\includegraphics[width=0.35\textwidth]{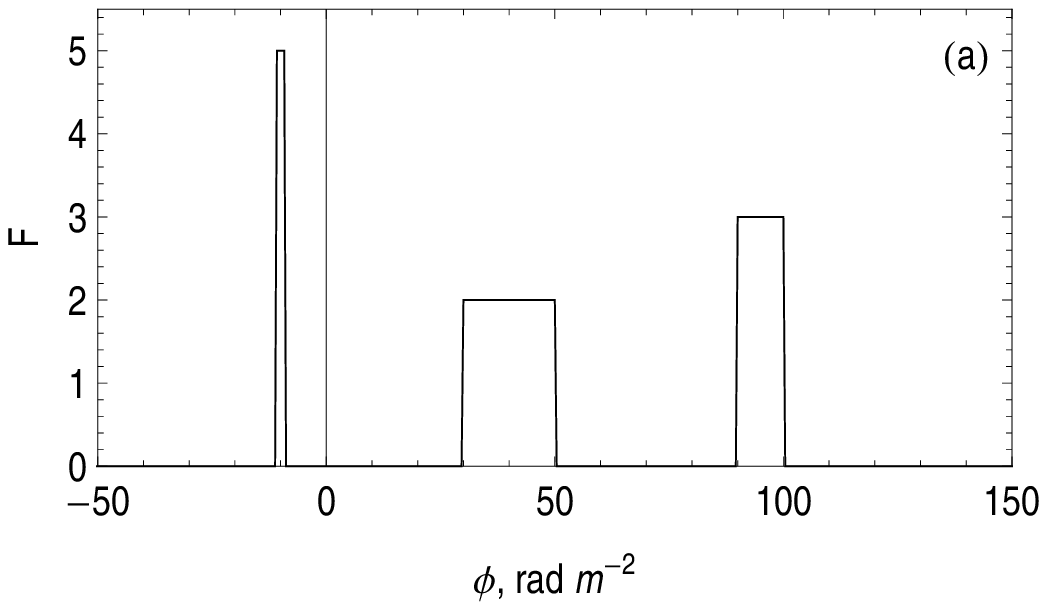}
\includegraphics[width=0.35\textwidth]{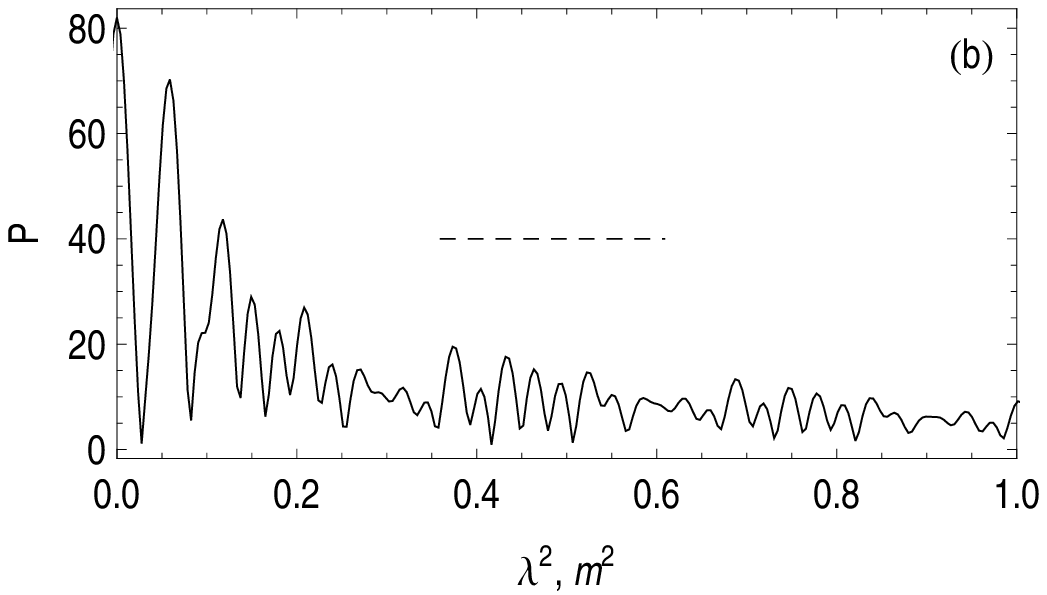}
\includegraphics[width=0.35\textwidth]{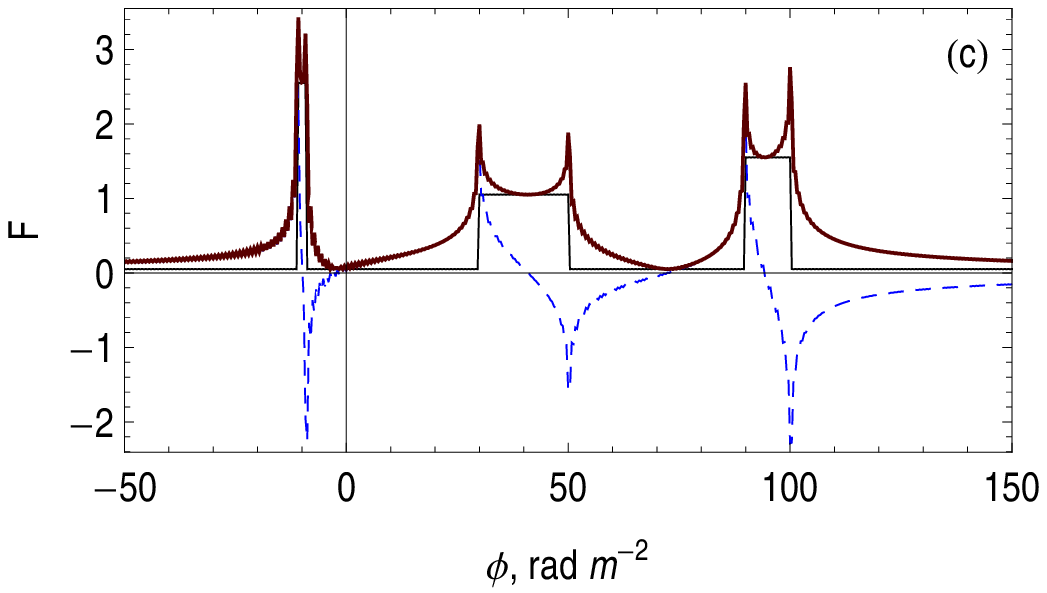}
\includegraphics[width=0.35\textwidth]{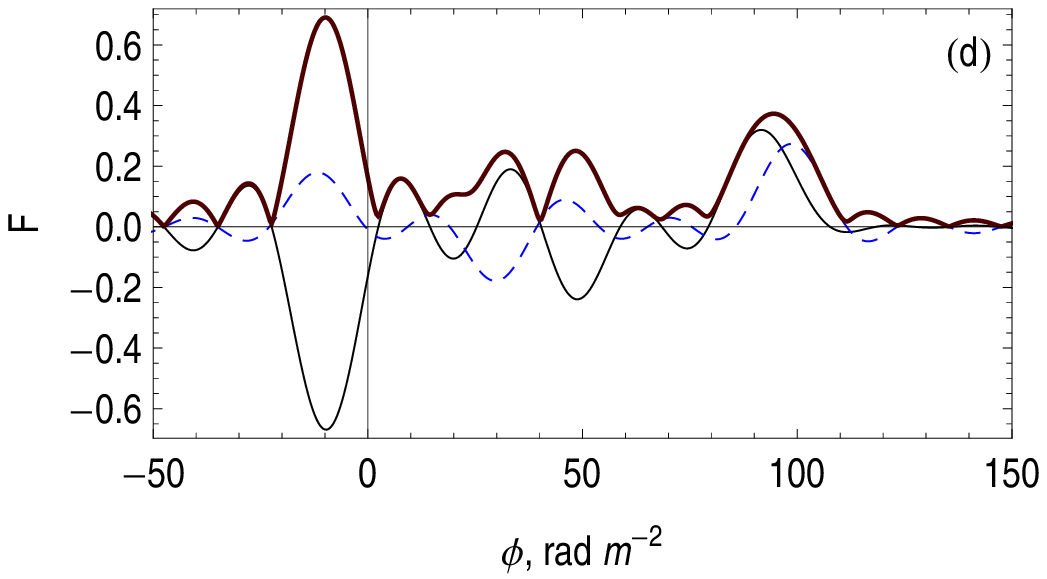}}
\caption {RM Synthesis reconstruction of a standard example from
\protect\cite{2005A&A...441.1217B}: (a) initial $F(\phi)$ which is
chosen purely real; (b) amplitude of $PI(\lambda^{2})$; (c)
$F(\phi)$  reconstructed with whole domain $\lambda^{2}>0$: real
part - thin solid, imaginary part - dashed, amplitude - thick solid;
(d) $F(\phi)$ reconstructed from the data of spectral band $ 0.6 <
\lambda < 0.78$\,m. The spectral window of observations is indicated
in panel (b) by horizontal dashed line.} \label{classex}
\end{figure}

Implementation of multichannel spectro-polarimetry on modern radio
telescopes provided observations of $P$ over a wide range of
$\lambda$ \citep[e.g.][]{2000A&A...356L..13H} which
made the use of Eq.~(\ref{Burn}) possible. This is the
idea of Faraday Rotation Measure Synthesis (RM
Synthesis) \citep{2005A&A...441.1217B} which opened
new perspectives in investigations of magnetic field of galaxies and
clusters of galaxies
\citep{2003A&A...403.1031H,2005A&A...441..931D,2008arXiv0804.4594B,2009arXiv0905.3995H}.

A key problem of RM Synthesis application is that $P$ is defined
only for $\lambda^2>0$ and in practice can be observed only in a
finite spectral band. Moreover, the maximum of $P$ in
practice can be located outside the available spectral band { (see e.g. Fig.~\ref{classex}b)}. Development of robust
methods for the reconstruction of $F$ from $P$ in a given spectral
range becomes crucial for the practical implementation of RM
Synthesis.

Fig.~\ref{classex} shows results of RM Synthesis applied to a
standard test as exploited by \cite{2005A&A...441.1217B}. Panel (a)
shows the function $F$, which includes three {\it real-valued}
box-like structures, panel (b) - the corresponding polarized
intensity $P$ (the dashed horizontal line shows the spectral window
$ 0.6 < \lambda < 0.78$\,m). We used a channel spacing of
$\delta\lambda=0.4$\,cm. Hereafter, $F$ and $P$ are numerically
evaluated in arbitrary but mutually consistent units. Note that $F$
is in general a complex-valued function. Its modulus defines the
emission and its phase defines the intrinsic position angle. Panel
(c) shows the result of the straightforward application of the RM
Synthesis algorithm to the physical range $\lambda^2 >0$, while
$P(\lambda^2)$ is set to zero for all negative $\lambda^2$. We see
that the real part of the reconstructed signal is the same as the
initial one (except that it has a twice lower amplitude), however,
the reconstructed signal obtains a substantial imaginary part with a
shape which is quite remote from the real part. This leads to a
change of the emission distribution and a loss of any information
concerning the position angle (apart from the central point of the
emission region, where the position angle correctly is zero). In the
context of chaotic magnetic fields in galaxy clusters this loss is
less important \citep{2005A&A...441..931D}, but in galactic magnetic
field studies it becomes crucial because the intrinsic position
angle determines the orientation of the regular magnetic field
component perpendicular to the line of sight. Fig.~\ref{classex}d
shows that the reconstruction becomes much more difficult if we
restrict the data to a relatively narrow spectral band $ 0.6 <
\lambda < 0.78$\,m. We see that even the sign of the reconstructed
real part can be wrong. In that case the algorithm for finite
spectral band
introduced by
\cite{2005A&A...441.1217B} was used.

A general message obtained from Fig.~\ref{classex} is that in order
to envisage possible ways to get a practical implementation of RM
Synthesis one has to include some additional information based on
the nature of the physical phenomena which provide the Faraday
rotation. Here we concentrate our efforts on the problems associated
with missing $P(\lambda^2)$ for $\lambda^2<0$.

\section{Improving the RM Synthesis algorithm}

The complex-valued intensity of polarized radio emission for a given
wavelength
\begin{equation}
\label{p-def}
P(\lambda^2) =  \int_{0}^{\infty} \varepsilon(z)e^{2i\chi(z)} e^{2i\phi(z) \lambda^2}  d z,
\end{equation}
is defined by the emissivity $\varepsilon$ and the intrinsic
position angle $\chi$  along the line of sight. Here $z$ is the
distance from observer to a point in the emitting region; the
integral is taken over the whole emitting region. If the Faraday
depth $\phi$ is a monotonic function of $z$ (which means
that $z$ is a single-valued function of $\phi$), we can define the
Faraday dispersion function as a function of Faraday depth
\begin{equation}
\label{F-to-z}
F(\phi)=\varepsilon(\phi)e^{2i\chi(\phi)}\left({d\phi}\over{dz}\right)^{-1}.
\end{equation}
In the ideal case, reconstructing the Faraday dispersion function
$F$ from (\ref{Burn}) and knowing the Faraday depth $\phi$ for any
$z$, one can derive the characteristics of radio emission
($\varepsilon$ and $\chi$) along the line of sight. They can be used
{ as a tomography}
in order to
derive some characteristics of the magnetic field distribution from
$F$. The  task of RM Synthesis is much more modest
and concerns the reconstruction of the Faraday function from the
observed polarized emission which itself is already a complicated
problem.

\begin{figure}\centering{
\includegraphics[width=0.35\textwidth]{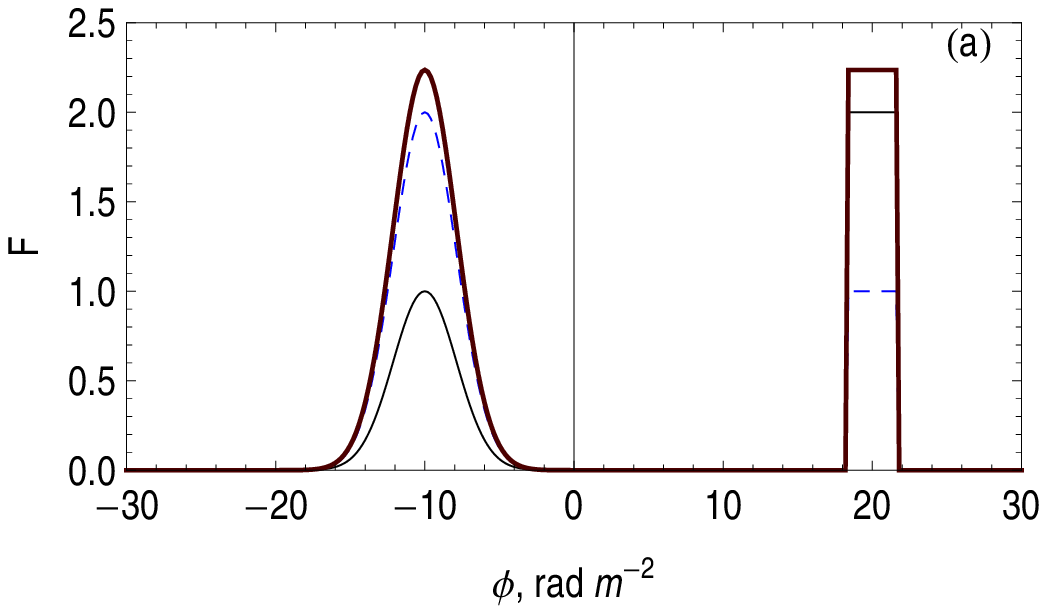}
\includegraphics[width=0.35\textwidth]{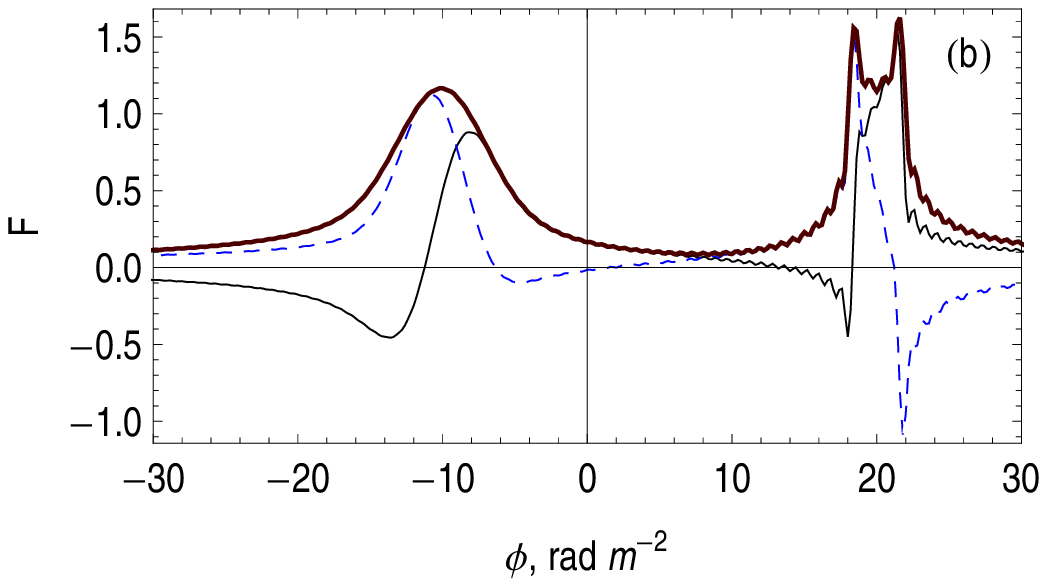}
\includegraphics[width=0.35\textwidth]{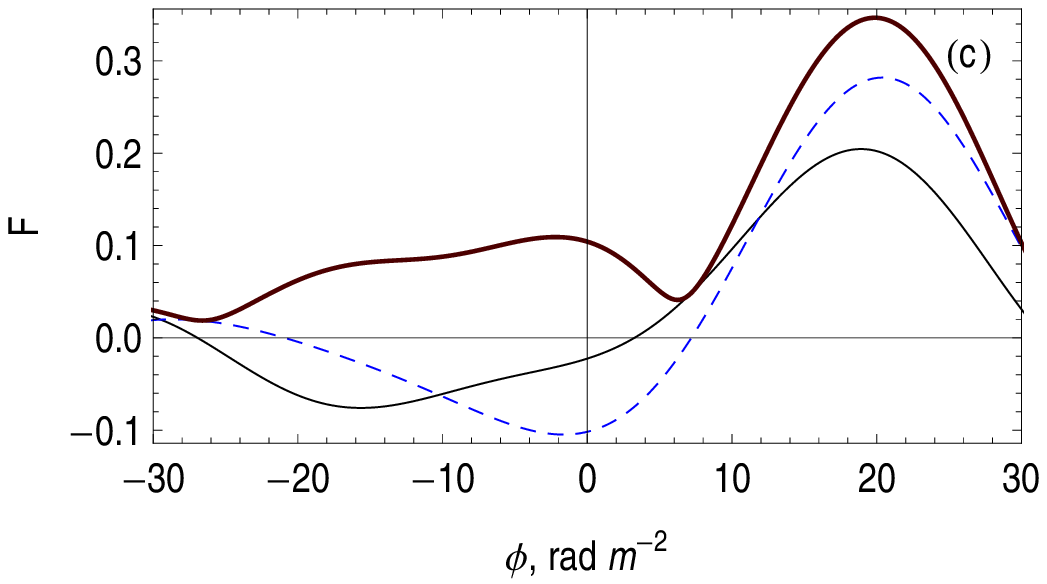}
\includegraphics[width=0.35\textwidth]{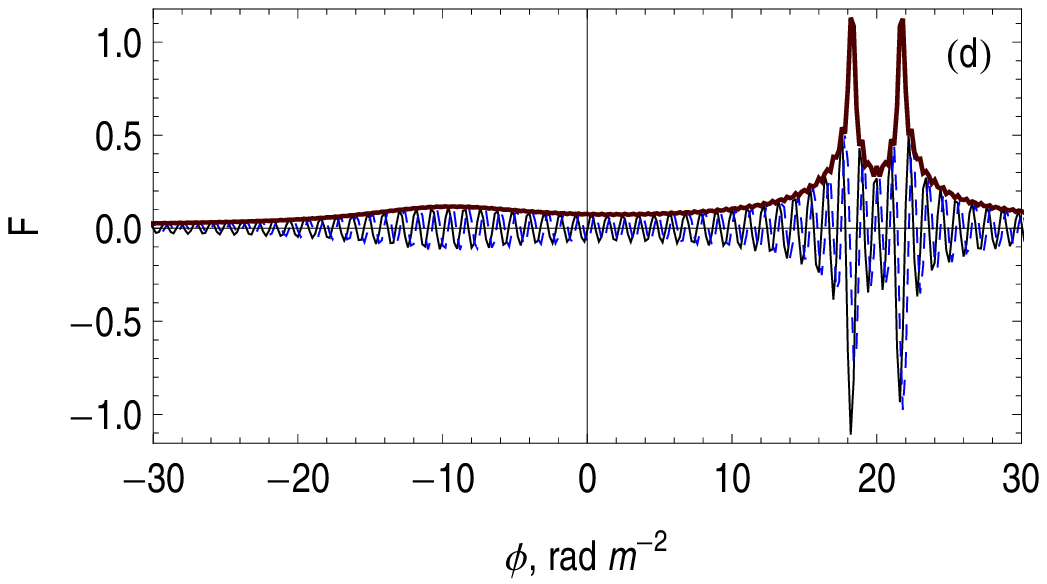}
} \caption {Standard RM Synthesis for a test Faraday dispersion
function. (a) Original test function which includes one
Gaussian and one box structure. Reconstructions: (b) using the whole
domain $\lambda^2>0$, (c) using the window $ 0.6 < \lambda <
0.78$\,m, (d) using the window $ 0.6 < \lambda < 2.5$\,m. Real part
- thin solid, imaginary part - dashed, amplitude - thick solid. }
\label{classex1}
\end{figure}

Let us consider a physically motivated simple example, i.e. $P$
produced by a two-layer system , to isolate and overcome the shortcomings of the RM
Synthesis technique. Each layer contains a homogeneous
magnetic field which has non-vanishing line-of-sight and
perpendicular components. Both layers are thought to be emitting and
rotating polarized radio waves. The corresponding $F(\phi)$ is shown
in Fig.~2a. It is important for the discussion below that the
analyzed signal has non-vanishing real and imaginary parts. The
absolute value of $F(\phi)$ indicates how much polarized emission
comes from a region with Faraday depth $\phi$ and its phase gives
the intrinsic position angle  (about $13^\circ$ and
$31^\circ$) of the emission. Just to illustrate the variety of
possible situations, we choose two different shapes of the slabs,
i.e. one slab with sharp boundaries and one with a Gaussian shape.

\begin{figure}\centering{
\includegraphics[width=0.35\textwidth]{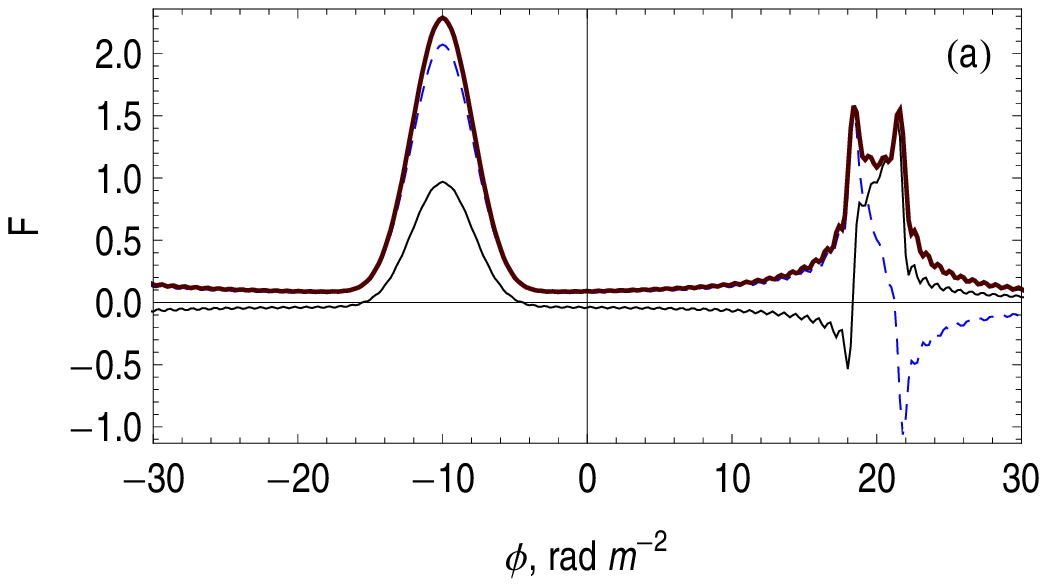}
\includegraphics[width=0.35\textwidth]{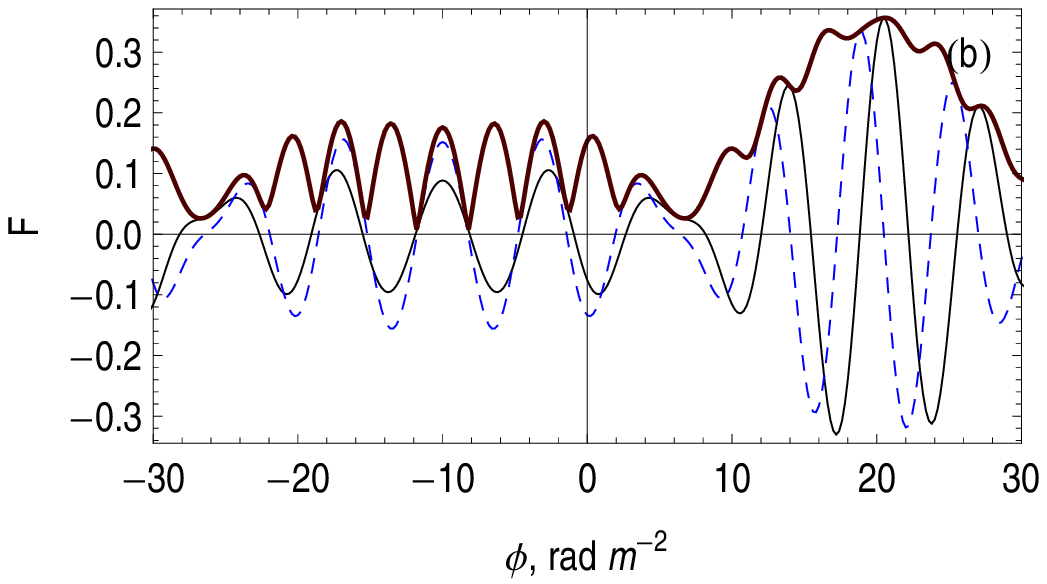}
\includegraphics[width=0.35\textwidth]{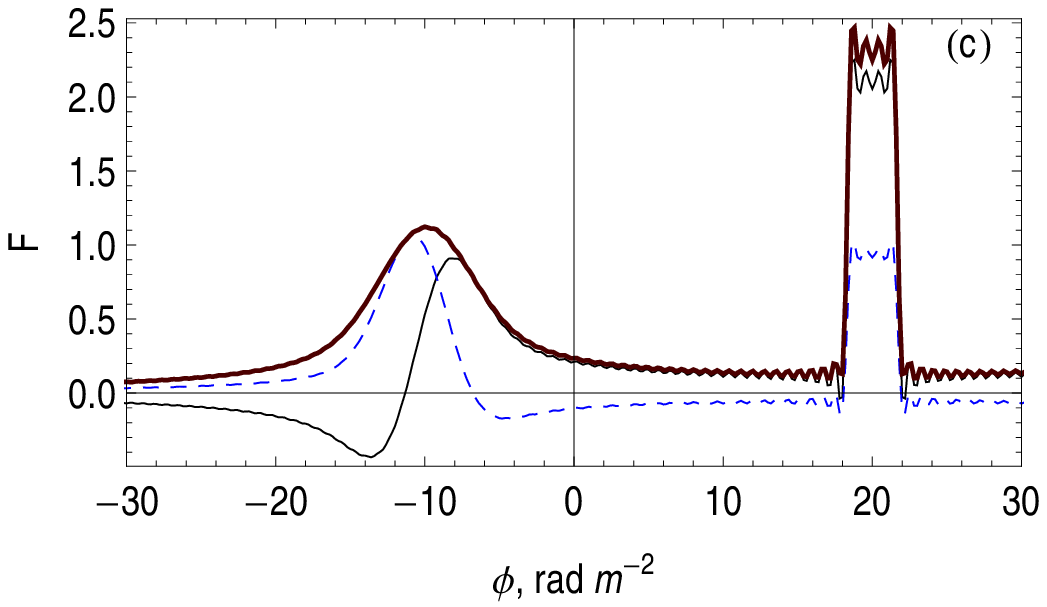}
\includegraphics[width=0.35\textwidth]{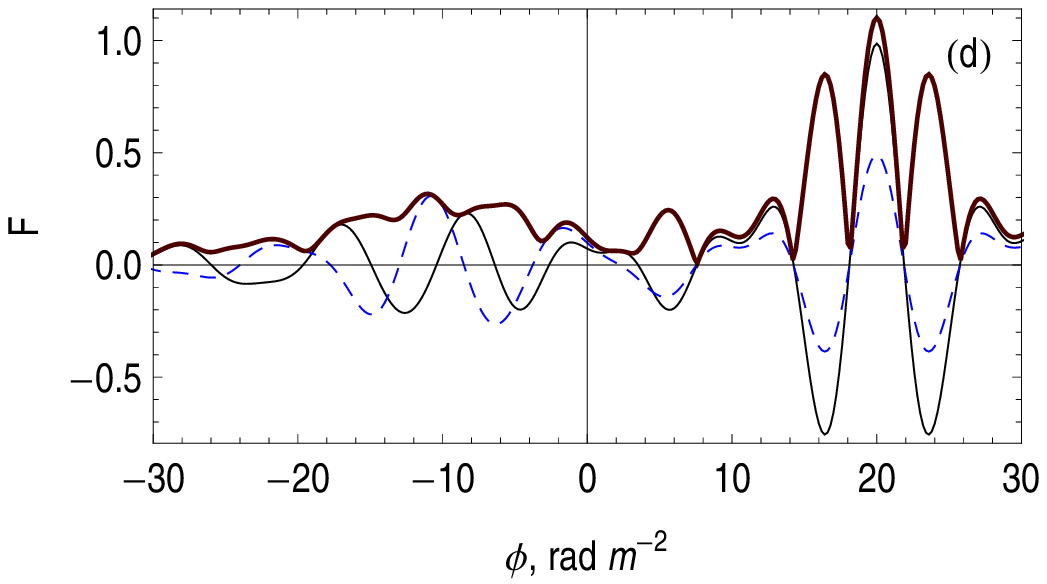}
} \caption {RM Synthesis for the test from
Fig.\protect{\ref{classex1}} using the extension of $P(\lambda^2)$
in the domain $\lambda^2<0$ defined by (\protect{\ref{cont}}). The
parameter $\phi_0$ is adjusted to the position of the left structure
(a,b) or right structure (c,d). The whole domain of $\lambda$ is
used in panels (a,c) and the spectral window $ 0.6 < \lambda <
0.78$\,m in panels (b,d). Real part - thin solid, imaginary part -
dashed, amplitude - thick solid.} \label{classex2}
\end{figure}

The result of the straightforward application of RM Synthesis where
the integral is taken over the physically admissible region
$\lambda^2>0$ is shown in Fig.~\ref{classex1}b. RM
Synthesis reproduces to some extent the absolute value of the
signal, but fails to reproduce its phase. A naive interpretation of
this result could be that field reversals occur in each layer, but
is obviously incorrect. In the same figure we show the result of
$F(\phi)$ reconstruction within the spectral band $ 0.6 < \lambda <
0.78$\,m (panel c). Then both structures become
diffuse with a more or less arbitrary phase. The last panel
illustrates what happens if the upper
wavelength boundary will be extended up to $\lambda =2.5$m
(as expected for the
Low Frequency Array (LOFAR) and the Square
Kilometre Array (SKA) telescopes). This extension essentially
improves the recognition of the sharp structure (the right one in
the figure) but almost does not affect the reconstruction of the
left (Gaussian) structure.

To avoid the non-uniqueness in the Faraday dispersion function
reconstruction, some additional information (or hypothesis) is
required. We suggest to improve the above reconstruction by
some constraint concerning the possible
symmetry of an isolated object.

Suppose that the expected objects are mainly galactic disks with
magnetic fields believed to be symmetric with respect to the
galactic equator. Then the desired $F$ should be even with respect
to the center of the given object. Therefore, we consider each
maximum of the reconstructed $F(\phi)$ separately and prescribe that
the continuation of $P(\lambda^2)$ to the region of $\lambda^2 <0$
has to be chosen in a way which makes $F(\phi)$ symmetric with
respect to the point $\phi = \phi_0$, where $\phi_0$ is the position
of the maximum under consideration. This means that
$F(2\phi_0-\phi)=F(\phi)$ and using the shift theorem one gets
\begin{equation}
P(-\lambda^2) = \exp (-4 i \phi_0 \lambda^2) P(\lambda^2).
\label{cont}
\end{equation}
The antisymmetric case can be considered as well with slight change
in the algorithm:  Eq.~(\ref{cont}) changes to
 $P(-\lambda^2) = -\exp (-4 i \phi_0 \lambda^2) P(\lambda^2)$.

Fig.~\ref{classex2} shows the results of reconstruction of the same
test but following the suggested continuation. The test function
includes two objects, while the algorithm includes only one
parameter $\phi_0$. Firstly, we performed the continuation adjusting
$\phi_0$ to the position of the left object (panel a). Then the
method gives realistic result for this object. The reconstructed
structure has no apparent internal field reversal and the ratio of
real and imaginary parts of $F(\phi)$, i.e. the phase, is correctly
reproduced. Position angles are restored with the
accuracy of $3^\circ$. Of course, the result for the other layer,
i.e. the second maximum of $|F(\phi)|$ in Fig.~\ref{classex2}
remains false. Panel (b) shows what happens if the range of
$\lambda$ covered by the observation is reduced to $ 0.6 < \lambda <
0.78$\,m. Instead of one peak one gets a sequence of peaks, which is
a usual result for a Fourier reconstruction using a narrow spectral
window. The suggested procedure does not suppress the sidelobes in
the standard Rotation Measure Spread Function (RMSF)
\citep{2009arXiv0905.3995H} but corrects the phase within the main
central peak. Of course, the amplitude of each peak is much less
than the amplitude of the peak in panel (a), however, the ratio of
real and imaginary parts of $F(\phi)$ in the central peak remains
realistic. If the parameter $\phi_0$ is chosen following the
position of the second object the method gives a correct
reconstruction for the right layer and fails to reproduce the left
one.
\begin{figure*}
\centering{
\includegraphics[width=0.33\textwidth]{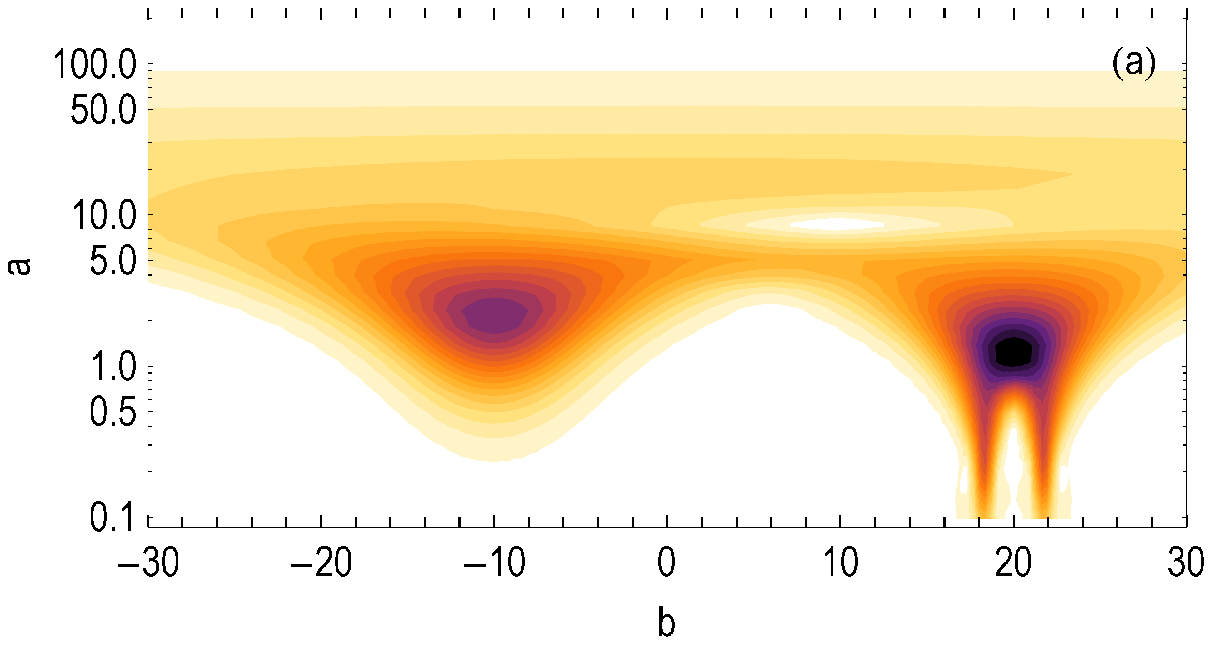}
\includegraphics[width=0.33\textwidth]{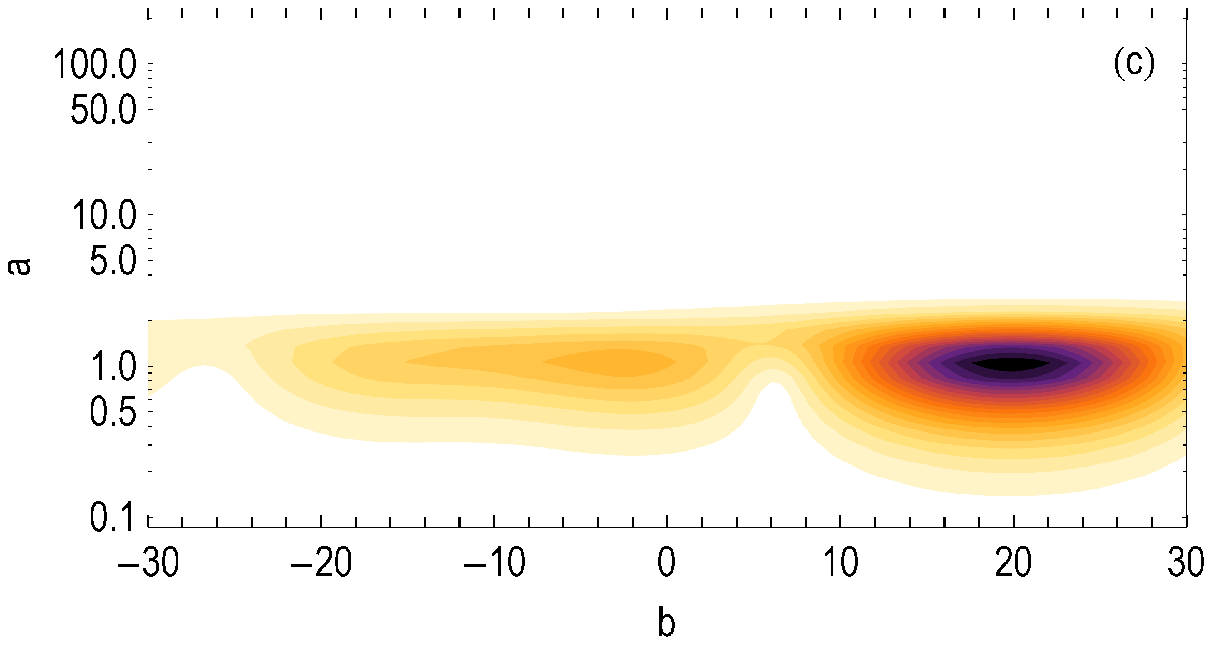}
\includegraphics[width=0.33\textwidth]{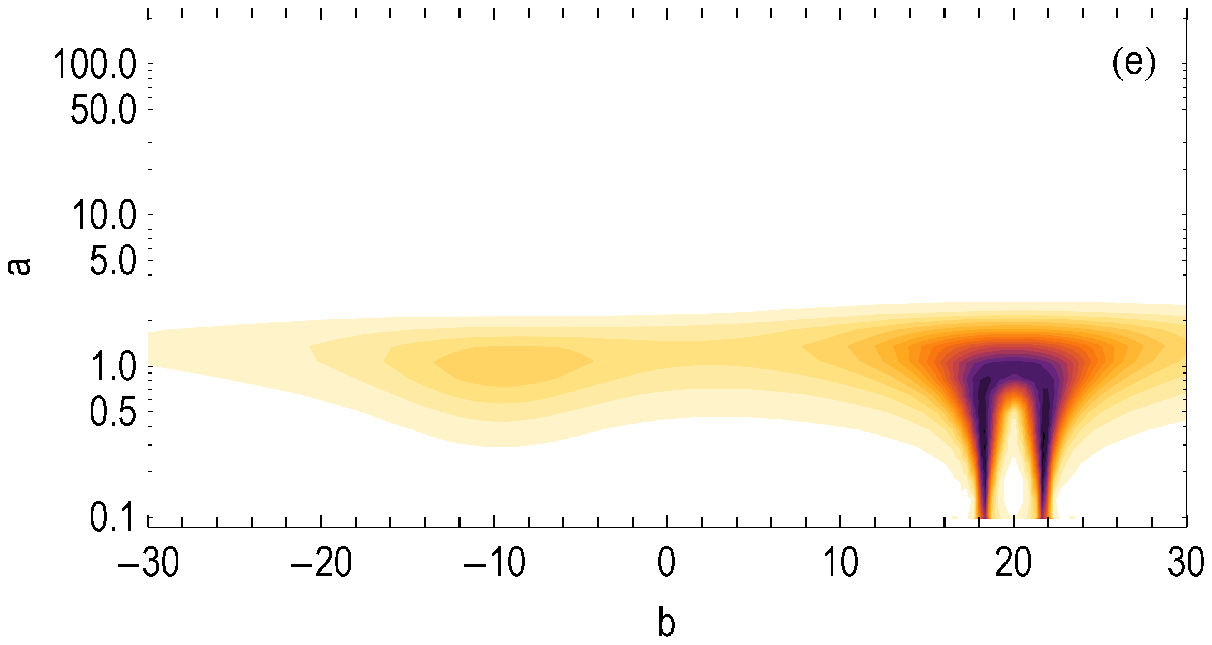}\\
\includegraphics[width=0.33\textwidth]{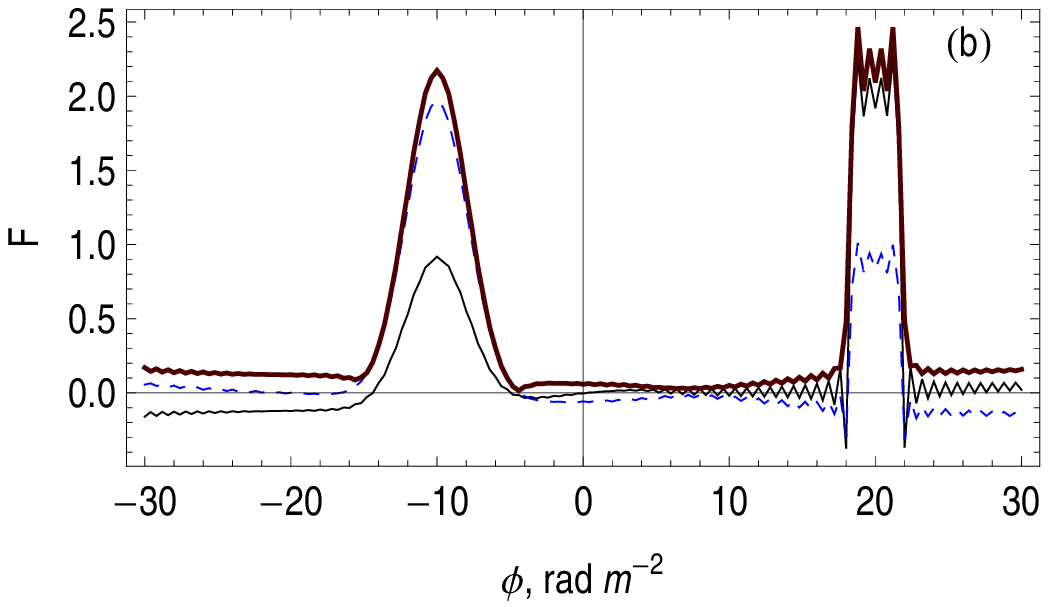}
\includegraphics[width=0.33\textwidth]{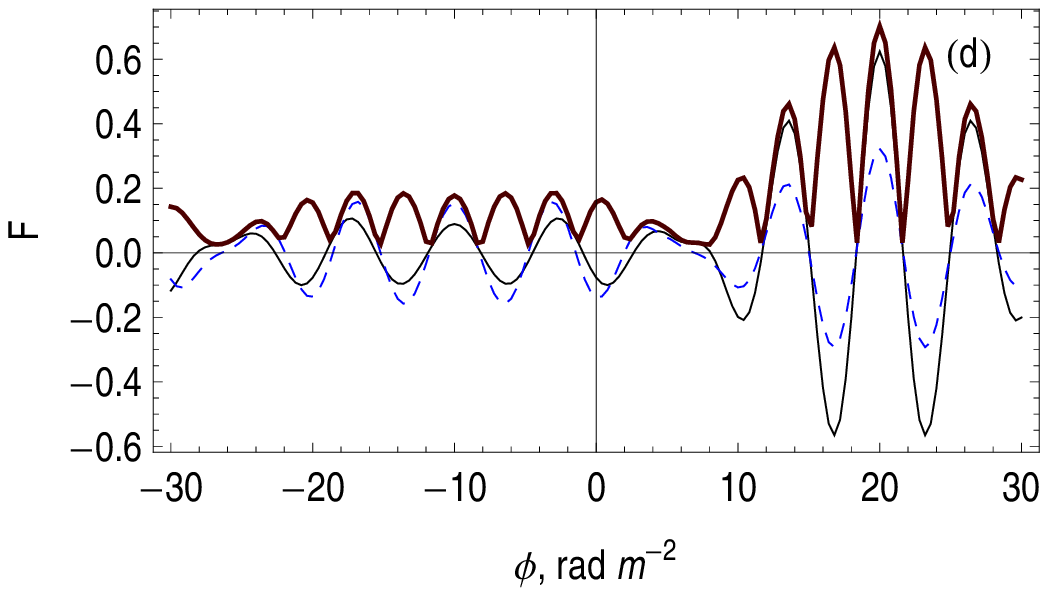}
\includegraphics[width=0.33\textwidth]{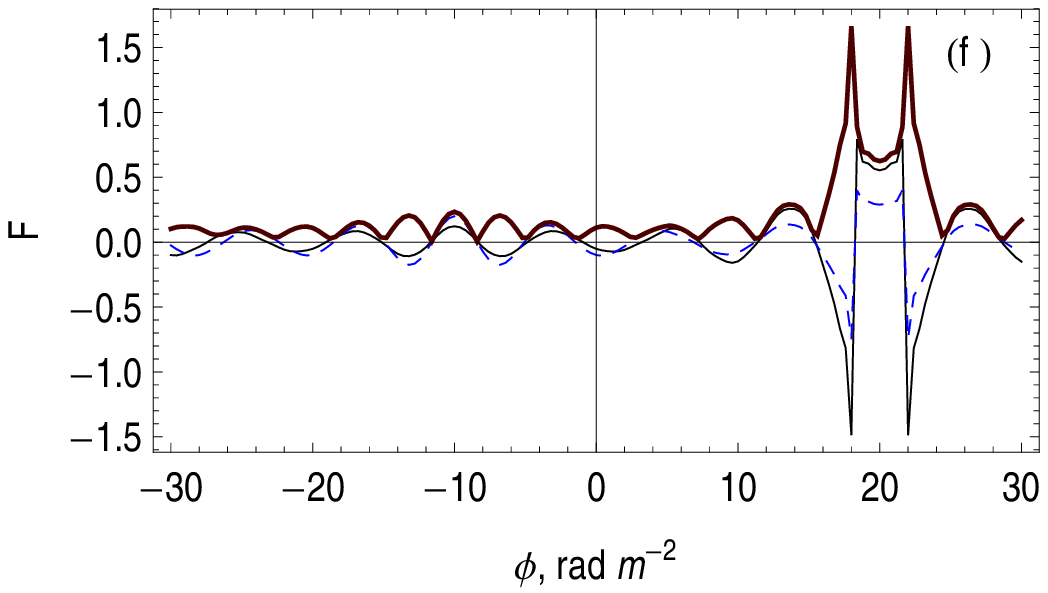}}
 \caption {Wavelet-based RM Synthesis for the test from
Fig.\protect{\ref{classex1}}. The modulus of wavelet coefficients on
the $(a,b)$ plane (panels a,c,e) and the result of reconstruction
(panels b,d,f) for whole domain of $\lambda$ (panels a,b) and the
windows $ 0.6 < \lambda < 0.78$\,m (panels c,d) and $ 0.6 < \lambda
< 2.5$\,m (panels e,f). Real part - thin solid, imaginary part -
dashed, amplitude - thick solid. } \label{classex3}
\end{figure*}
An obvious shortcoming of the method exploited is its local nature:
We obtain a realistic shape of a chosen maximum and ignore what
happens with the other one. A natural extension is
to apply the recommendation of Eq.~(\ref{cont}) locally to each
maximum. This extension brings the idea of
wavelets into consideration.

\section{RM Synthesis and wavelets}

Wavelet transform presents a kind of ``local'' Fourier transform,
allowing us to isolate a given structure in physical space and the
Fourier space. Let us define the wavelet transform of the Faraday
dispersion function $F(\phi)$ as
\begin{equation}
\label{wF_d}
w_F(a,b) = {{1}\over {|a|}} \int\limits_{ - \infty }^\infty
{F(\phi)\psi ^\ast \left( {\frac{\phi - b}{a}} \right)d\phi} ,
\end{equation}
where $\psi(\phi)$ is the analyzing wavelet, $a$ defines the scale
and $b$ defines the position of the wavelet. Then the coefficient
$w_F$ gives the contribution of corresponding structure into the
function $F$.

The function $F$ can be reconstructed using the inverse transform
(see, e.g. \cite{1992ApMat..61.....D})
\begin{equation}
\label{wF_inv}
F(\phi) = \frac{1}{C_\psi }\int\limits_{-\infty}^\infty {\int\limits_{ -
\infty }^\infty {\psi \left( {\frac{\phi - b}{a}} \right)w_F\left(
{a,b} \right)\frac{d a \, d b}{a^2}} }.
\end{equation}
The reconstruction formula (\ref{wF_inv}) exists under condition that
 \begin{equation}
\label{adm}
C_\psi = \frac1{2}\int\limits_{ - \infty }^\infty
{\frac{\vert \hat {\psi }(k )\vert ^2}{ |k| }d k < \infty } .
\end{equation}
Here $\hat{\psi }(k ) = \int {\psi (\phi)e^{ - ik\phi}d\phi} $ is
the Fourier transform of the analyzing wavelet $\psi(\phi)$.

Let us emphasize that the inverse formula (\ref{wF_inv}) is usually
written for real signals. Then the scale parameter $a$ is positively
defined and the integral is taken for $0< a<\infty$. In the case
of a complex-valued function, the range of $a$ can be limited by
positive values $a>0$ by taking a {\it real} analyzing wavelet
$\psi(x)$. In general case of a complex-valued function and a complex
wavelet, the scale parameter $a$ should be extended into the domain
of negative values (like wave numbers in Fourier space).

For the sake of definiteness, we use as the analyzing wavelet the
so-called Mexican hat $\psi (\phi) = (1-\phi^2) \exp (-\phi^2/2)$.
The wavelet is real, however, the function $P$ is complex, so that
the wavelet coefficients $w_F$ are complex as well. For the chosen
wavelet $w_F(-a,b)=w_F(a,b)$ and $C_\psi = 1$.

Using the definition of the wavelet transform (\ref{wF_d})
and  relation (\ref{Burn}) we can directly define the
wavelet decomposition of the Faraday dispersion function from the
polarized intensity $P(\lambda^2)$
\begin{equation}
\label{wF_P} w_F(a,b) = {{1}\over {\pi}} \int\limits_{ -
\infty}^\infty {P(\lambda^2)e^{-2ib \lambda^2} \hat{\psi}^\ast
\left( -2a \lambda^2 \right)d \lambda^2} .
\end{equation}

Note that in the case of real $F$
the problem of negative
$\lambda^2$ can be solved using progressive wavelets, whose Fourier
image is localized in the domain of positive wave numbers. Thus
using this kind of wavelets one avoids the problem of the
$P(\lambda^2)$ continuation in the domain $\lambda^2 <0$.

For the general case, we divide Eq.~(\ref{wF_P}) in two parts $w_F(a,b) =w_-(a,b)+w_+(a,b)$, where
\begin{eqnarray}
\label{wF_P2} w_-(a,b)&=& {{1}\over {\pi}}
\int\limits_{ - \infty }^0
{P(\lambda^2)e^{-2ib \lambda^2} \hat{\psi}^\ast \left( -2a
\lambda^2 \right)d\lambda^2},  \\   w_+(a,b)&=&{{1}\over {\pi}} \int\limits_0^\infty
{P(\lambda^2)e^{-2ib \lambda^2} \hat{\psi}^\ast \left( -2a \lambda^2 \right)d\lambda^2}.
\end{eqnarray}

We propose the following algorithm: Firstly, knowing $P(\lambda^2)$
for $\lambda^2>0$ we calculate the coefficients $w_+(a,b)$ and we
recognize the dominating structures in the map $|w_+(a,b)|$. The
coordinate $b$ of the corresponding maximum gives us the value of
$\phi_0^i$, where upper index $i$ indicates the number of the
structure. Then we reconstruct the coefficients $w_-(a,b)$ following
the idea of Eq.~(\ref{cont}), but reformulated for the local domain
in wavelet space $(a,b)$. Namely, we define
\begin{equation}
\label{w-w+}
w_-(a,b)=w_+\left(a,2\phi_0^i(a,b)-b\right),
\end{equation}
where the parameter $\phi_0^i(a,b)$ for the given point $(a,b)$ is chosen
according to the structure $i$ which dominates in its vicinity.

\begin{figure}\centering{
\includegraphics[width=0.35\textwidth]{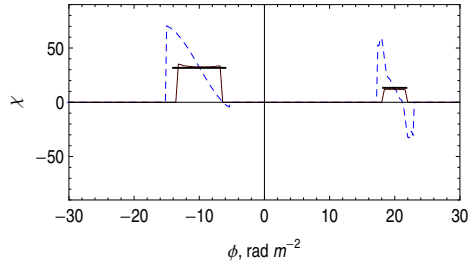}
} \caption {The intrinsic position angle $\chi$ for the test from
Fig.\protect{\ref{classex1}}(a) has been reconstructed with standard (dashed) and wavelet-based (thin solid) RM Synthesis. The whole domain of $\lambda$ is
used. Thick solid lines show initial $\chi$ in the location of the both structures.} \label{comp}
\end{figure}

Now we apply the suggested algorithm to the test function from
Fig.~\ref{classex1}. The map $|w_+(a,b)|$ presented in
Fig.~\ref{classex3}a demonstrates two well-defined structures. The
$b$-coordinates of the maxima are taken as $\phi_0^i$. The
result of the reconstruction (see Fig.~ \ref{classex3}b) shows that
the method reproduces the amplitude and phase of $F(\phi)$ for both layers.  The reconstruction here is performed using
$P(\lambda^2)$ for the whole range $\lambda^2>0$. { The comparison of the reconstructed position angle using standard and wavelet-base RM Synthesis is shown in Fig.~\ref{comp}. The suggested algorithm gives correct value for $\chi$ within both emission regions.}
Panels (c,d) show
what happens for the reconstruction using the spectral window $ 0.6
< \lambda < 0.78$\,m. One can see the wavelet map is empty in
its substantial part $a>2$, however, the structures
remain well-recognizable (panel c). The reconstructed $F$ contains
several oscillations in domains related to both layers. The
amplitude of each oscillation becomes much lower than that in panel
(b), however, the ratio of the real and imaginary parts in the
central maxima remain correct. The third couple of panels shows the
reconstruction within the extended window $ 0.6 < \lambda < 2.5$\,m.
This extension allows one to keep the horn-like structures in the
bottom of the wavelet plane (panel e) which provide the
reconstruction of sharp boundaries of the box-like structure (panel
f).

\section{Conclusions}

The development of multi-channel observations of
polarized radio emission opens promising perspectives in the
understanding of cosmic magnetic fields on galactic and
intergalactic scales. The first fruitful applications
of RM Synthesis suggested in this context include
the recognition of local structures in the Milky Way
\citep{2003A&A...403.1031H}, clusters of galaxies
\citep{2005A&A...441..931D} and spiral galaxies
\citep{2009arXiv0905.3995H}. However, in general the RM Synthesis
algorithm contains a fundamental problem emerging from the fact that
the reconstruction formula requires the definition of complex
polarized intensity in the range $-\infty <\lambda^2 < \infty$. In
this paper we introduce a simple method for continuation of observed
complex polarized intensity $P(\lambda^2)$ into the domain of
negative $\lambda^2<0$. The method is suggested in context of
magnetic field recognition in galactic disks, for which the magnetic
field strength is supposed to have a maximum in the equatorial
plane.

The suggested method is quite simple when applied
to a single structure on the line of sight. Recognition
of several structures on the same line of sight requires a more
sophisticated technique. The problem of structure separation is
resolved using the wavelet decomposition. A simple test example
demonstrates the applicability of this method. { The polarization angle reconstruction is significantly improved over the standard technique.}
The wavelets can be useful to also overcome some
other problems of RM Synthesis, related to the multi-band structure
of the observational domain in $\lambda$-space, noise filtration,
etc \citep[e.g.][]{1997ApJ...483..426F,2001MNRAS.327.1145F}.
The method essentially improves the possibilities for
reconstruction of complicated Faraday structures using the
capabilities of modern radio telescopes.

Finally note that our simple examples illustrate that the extension
of the observational band into the long-wavelength domain is
helpful for the recognition of structures with sharp
boundaries, while the short-wavelength domain is crucial for
the reconstruction of smooth structures.

\label{lastpage}

\section*{Acknowledgments} This work was supported by the
DFG-RFBR grant 08-02-92881.

\bibliographystyle{mn2e} 
\bibliography{ref}

\end{document}